# Why semi-classical electrodynamics is not gauge invariant

## A M Stewart

Research School of Physical Sciences and Engineering,
The Australian National University,
Canberra ACT 0200, Australia.

### Abstract

It is shown that in semi-classical electrodynamics, which describes how electrically charged particles move according to the laws of quantum mechanics under the influence of a prescribed classical electromagnetic field, only a restricted class of gauge transformations is allowed. This lack of full gauge invariance, in contrast to the situation in classical and quantum electrodynamics which are fully gauge invariant theories, is due to the requirement that the scalar potential in the Hamiltonian of wave mechanics represent a physical potential. Probability amplitudes and energy differences are independent of gauge within this restricted class of gauge transformation.

**1.     Introduction**
This paper is about the role of gauge in the wave mechanics of a charged particle moving non-relativistically in an externally prescribed electromagnetic field. The motion of the charged particle is described by the free particle Schrodinger equation supplemented by minimal coupling of the charge to the potentials of the electromagnetic field.

The question of the effect of gauge transformations on the Hamiltonian of the Schrodinger equation and on transition probability amplitudes has had a long and sometimes contentious history. It was suggested by Lamb (1952) that in the hydrogenic spectrum different line shapes would be calculated depending on whether the perturbation between the electron and the electromagnetic field was taken to be ***A.p*** or ***E.r***, the different forms of the interaction being related by a gauge transformation (Stewart 2000). Although the particular case raised by Lamb appears to have been resolved subsequently by considering the intermediate states in the radiative process, see Cohen-Tannoudji, Dupont-Roc *et al.* (1989) for a review (but see also the discussion of Woolley 2000), the suggestion remained that transition amplitudes and probabilities could depend on gauge. It was also pointed out by Yang (1976) that the Hamiltonian of the problem was not independent of gauge and so a question arose as to the interpretation of the energy. There was much discussion of the issue in the literature (Aharonov and Au 1981, Aharonov and Au 1983, Feuchtwang, Kazes *et al.* 1984a, Feuchtwang, Kazes *et al.* 1982, Feuchtwang, Kazes *et al.* 1984b, Haller 1984, Kobe 1978, Kobe 1982, Kobe 1984a, Kobe 1984b, Kobe and Yang 1987a, Kobe and Yang 1987b, Power 1989, Power and Thirunamachandran 1978, Rau 1996, Schlicher, Becker *et al.* 1984, Yang 1981, Woolley 2000) but no clear consensus appears to have emerged.

This paper proposes a new approach to the problem in which gauge freedom is explicitly maintained throughout by the presence of an arbitrary gauge function in the equations. In section **2** we recall the derivation of the Schrodinger equation for a charged particle in an electromagnetic field and in section **3** we obtain the basis functions that are used to describe stationary states and that are needed for use in perturbation theory when an arbitrary gauge is adopted. In section **4** we obtain matrix elements of operators in an arbitrary gauge and find that the Hamiltonian is not gauge invariant. In section **5** we realize that for the scalar potential to represent the physical potential a restricted set of gauges must be used, namely those that are a sum of functions that depend respectively on spatial coordinates only and of time only. In section **6** we find that probability amplitudes are essentially independent of gauge within this restricted set and in section **7** we discuss why semi-classical electrodynamics, in contrast to classical and quantum electrodynamics, is not fully gauge invariant. In the two appendices a gauge transformation needed for the basis functions describing time-independent fields and a result concerning the canonical commutation relations are obtained.

## 2. Schrodinger equation

The Lagrangian of a particle of charge $e$ and mass $m$ at position $\mathbf{r}$ and time $t$ in the electromagnetic field described by vector potential $\mathbf{A}(\mathbf{r},t)$ and scalar potential $\phi(\mathbf{r},t)$ is taken to be (Doughty 1980, Schwinger, DeRaad Jr. *et al.* 1998)

$$L = mv^2/2 + e(\mathbf{v}\cdot\mathbf{A} - \phi) \quad , \qquad (1)$$

where the velocity $\mathbf{v}$ is given by $\mathbf{v} = d\mathbf{r}/dt$.

If the electromagnetic potentials $\mathbf{A}$ and $\phi$ are transformed to $\mathbf{A}_\chi$ and $\phi_\chi$ by

$$\mathbf{A} \rightarrow \mathbf{A}_\chi = \mathbf{A} + \nabla\chi \quad \text{and} \quad \phi \rightarrow \phi_\chi = \phi - \partial\chi/\partial t \quad , \qquad (2)$$

where $\nabla$ is the gradient operator with respect to $\mathbf{r}$ and the subscript attached to the potentials indicates the gauge function that is used, then the electromagnetic fields $\mathbf{B}$ and $\mathbf{E}$ given by

$$\mathbf{B} = \nabla\times\mathbf{A} \quad \text{and} \quad \mathbf{E} = -\nabla\phi - \partial\mathbf{A}/\partial t \qquad (3)$$

are unchanged. The arbitrary scalar field $\chi(\mathbf{r},t)$ is known as the gauge function and the transformation as a gauge transformation. The gauge function is required to satisfy the condition $\{(\partial/\partial i)(\partial/\partial j) - (\partial/\partial j)(\partial/\partial i)\}\chi = 0$ everywhere where $i$ and $j$ are any pair of the coordinates $x, y, z$ and $t$. Under the gauge transformation, the Lagrangian in the new gauge becomes

$$L_\chi = mv^2/2 + e\mathbf{v}\cdot(\mathbf{A} + \nabla\chi) - e(\phi - \partial\chi/\partial t) \quad , \qquad (4)$$

the additional terms amounting to the total derivative $d\chi/dt$ thereby leaving the Euler-Lagrange equation

$$m\,d\mathbf{v}/dt = e(\mathbf{E} + \mathbf{v}\times\mathbf{B}) \qquad (5)$$

unchanged. In the new gauge the canonical momentum of the particle $\mathbf{p}_\chi = \partial L_\chi/\partial \mathbf{v}$ is given by

$$\mathbf{p}_\chi = m\mathbf{v} + e(\mathbf{A} + \nabla\chi) \qquad (6)$$

and the Hamiltonian $H_\chi(\mathbf{p}_\chi,\mathbf{r},t) = \mathbf{v}\cdot\mathbf{p}_\chi - L_\chi$ by

$$H_\chi(\mathbf{p}_\chi,\mathbf{r},t) = \{\mathbf{p}_\chi - e(\mathbf{A} + \nabla\chi)\}^2/2m + e(\phi - \partial\chi/\partial t) \quad . \qquad (7)$$

The passage to the wave mechanics of Schrodinger is made by changing classical observables into operators and taking the Hamiltonian to be equivalent to the operator $i\hbar(\partial/\partial t)$ acting upon a wavefunction $\Psi(\mathbf{r},t)$ which is the probability density amplitude for the particle to be at position $\mathbf{r}$ at time $t$ and $\hbar$ is Planck's constant. Further, from the canonical commutation relations $[r^i, p_\chi^j] = i\hbar\delta_{i,j}$ of Dirac (1947) the canonical momentum $\mathbf{p}_\chi$ becomes the differential operator $-i\hbar\nabla$ (but see Appendix **1**). This, true in all gauges, leads to the Schrodinger wave equation

$$H_\chi(-i\hbar\nabla,\mathbf{r},t)\Psi_\chi(\mathbf{r},t) = i\hbar(\partial/\partial t)\Psi_\chi(\mathbf{r},t) \qquad , \qquad (8)$$

the Hamiltonian $H_\chi$ in the gauge $\chi$ being given by (7).

From the readily verified identity

$$\{(\mathbf{p} - e(\mathbf{A} + \nabla\chi))\}^s \Psi \exp\{ie\chi/\hbar\} = \exp\{ie\chi/\hbar\}\{\mathbf{p} - e\mathbf{A}\}^s \Psi \qquad , \qquad (9)$$

where $\mathbf{p} = -i\hbar\nabla$, $s$ is a positive integer and $\Psi$ is any function of $\mathbf{r}$ and $t$, it follows that if, under a gauge transformation, the wavefunction is assumed to transform according to

$$\Psi_0(\mathbf{r},t) \to \Psi_\chi(\mathbf{r},t) = \Psi_0(\mathbf{r},t)\exp\{ie\chi(\mathbf{r},t)/\hbar\} \qquad , \qquad (10)$$

then $H_\chi\Psi_\chi = i\hbar(\partial/\partial t)\Psi_\chi$ implies $H_0\Psi_0 = i\hbar(\partial/\partial t)\Psi_0$ or, in other words, the Schrodinger equation is invariant in form under a gauge transformation.

### 3.   Construction of basis functions

The object of wave mechanics is to obtain $\Psi$ from equation (8) given a particular set of potentials, and obtain physical quantities from the expectation values of the appropriate operators. When equation (8) admits of no simple solution the usual method of solving it is to expand $\Psi(\mathbf{r},t)$ in a set of basis functions $\Psi_n(\mathbf{r},t)$ that are solutions of some simpler equation as

$$\Psi_\chi(\mathbf{r},t) = \sum_n a_{\chi,n}(t)\, \Psi_{\chi,n}(\mathbf{r},t) \qquad , \qquad (11)$$

where the $a_{\chi,n}(t)$ form a set of expansion coefficients. The quantum numbers $n$ are associated with a particular quantum state of the simpler system and the subscript $\chi$ is included to indicate that all quantities may in general depend on the gauge that is used. The probability amplitude $p_{\chi,n}'(t)$ for the system to be in state $n$ at time $t$ is defined to be the projection of $\Psi_{\chi,n}$ onto $\Psi_\chi$

$$p_{\chi,n}'(t) = \int \Psi_{\chi,n}^*(\mathbf{r},t)\Psi_\chi(\mathbf{r},t)\mathrm{d}\mathbf{r} \qquad , \qquad (12)$$

where d$\mathbf{r}$ is the three-volume element at $\mathbf{r}$. If the $\Psi_{\chi,n}(\mathbf{r},t)$ are orthonormal with respect to integration over space it follows that the probability amplitude is equal to $a_{\chi,n}(t)$, depending, in general, on gauge. The probabilities themselves $|a_{\chi,n}(t)|^2$, being observables, should not depend on gauge but the amplitudes, on the basis of this argument, *might* depend on gauge.

Equation (8) cannot be solved by separation of variables to obtain the simpler set of functions unless the operator on the left, $H_\chi$, is independent of time. If the fields $\mathbf{E}(\mathbf{r})$ and $\mathbf{B}(\mathbf{r})$ have no dependence on time it is always possible to find a gauge in which the potentials have no time dependence either; we call them $A^0(\mathbf{r})$ and $\phi^0(\mathbf{r})$. This claim is justified by the following argument. The vector potential $A$ can, as can any three-vector, be expressed as the sum of a solenoidal part $A_T$, with $\nabla \cdot A_T = 0$, and a longitudinal part $\nabla \alpha(r,t)$ (Panofsky and Philips 1955). By making a gauge transformation with gauge function equal to $-\alpha(r,t) + \beta(r) - \gamma t$, where $\gamma$ is a number, the longitudinal part becomes $\nabla \beta(r)$, independent of time. From $\partial B/\partial t = 0$ it follows that the relation $\partial(\nabla \times A_T)/\partial t = 0$ must hold. The useful solution to this equation is $\partial A_T/\partial t = 0$, the other solution $\nabla \times A_T = 0$ is uninteresting as it gives zero $B$. Accordingly, $A^0(r) = A_T(r) + \nabla \beta(r)$ and is a function of $r$ alone. The electric field is given in terms of the transformed potentials by $E = -\nabla \phi^0 - \partial A^0/\partial t$. Since $\partial E/\partial t = 0$ it follows that $\partial \nabla \phi^0/\partial t = 0$; again the useful solution is $\partial \phi^0/\partial t = 0$, the other solution $\nabla \phi^0 = 0$ is also uninteresting as it gives zero $E$. Accordingly, $\phi^0$ as well as $A^0$ can be chosen to be a function of $r$ only. We note that other time independent gauges $A^{0\prime} = A^0 + \nabla f(r)$ and $\phi^{0\prime} = \phi^0 - g$ can be constructed with a gauge transformation using a gauge function $\chi^0(r,t) = f(r) + gt$, where $g$ is a number. Of course, time dependent gauges for these time independent fields may be constructed by using a gauge function that depends more generally on time, but are not appropriate for the present purpose.

We give two examples of time independent gauges. The first is the potentials of electro- and magnetostatics expressed in terms of the time independent charge $\rho(r)$ and current $J(r)$ densities (Panofsky and Philips 1955)

$$\phi_1^0(r) = \frac{1}{4\pi\varepsilon_0} \int \frac{\rho(r')\,dr'}{|r-r'|} \quad \text{and} \quad A_1^0(r) = \frac{\mu_0}{4\pi} \int \frac{J(r')\,dr'}{|r-r'|} \quad . \tag{13}$$

In the second the potentials are expressed in terms of the fields in the multipolar gauge (Kobe 1982, Stewart 1999, Valatin 1954, Woolley 1973)

$$\phi_2^0(r) = -(r-R)\cdot\int_0^1 E(q)\,du \quad \text{and} \quad A_2^0(r) = -(r-R)\times\int_0^1 B(q)u\,du \quad , \tag{14}$$

where $q = ur + (1-u)R$. The multipolar gauge describes generally the potentials that arise from fields that depend on time but of course can be used equally well with fields that do not. It is shown in Appendix **2** that the two sets of potentials above are related by a gauge transformation with gauge function $\chi^0(r,t) = f(r) + gt$, and $f(r)$ and $g$ are obtained explicitly. Of course, full gauge arbitrariness remains in expressions (13) and (14) because another arbitrary gauge function $\chi'$ can always be added to them according to equation (2). The gauge must then be fixed by setting $\chi'$ to zero, but it is only legitimate to do this if it can be demonstrated that the result of the calculation of any physical property is independent of $\chi'$.

However, due to the presence of $\chi$ the Hamiltonian operator on the left of equation (8) is still time dependent even when the potentials are static. In this circumstance the wavefunction cannot be separated into the product of a time dependent part and a space dependent part. To find a solution to equation (8) with static potentials we temporarily set the gauge function to zero. This gives

$$H_0^0 \Psi_0(r,t) = \{(p - eA^0)^2/2m + e\phi^0\}\Psi_0(r,t) = i\hbar(\partial/\partial t)\Psi_0(r,t) \quad , \tag{15}$$

the zero superscript indicating that the potentials are time independent, the zero subscript indicating that the gauge function is set to zero. The operator on the left-hand side is now independent of time so the wavefunction may be factored into a space-dependent part $\psi(r)$ and a time-dependent part. These separate to give

$$\Psi_{0,n}(r,t) = \psi_n(r)\exp(-iE_n t/\hbar) \qquad , \qquad (16)$$

with $E_n$ and $\psi_n(r)$ given by the eigenvalue equation

$$\{(p - eA^0)^2/2m + e\phi^0\}\psi_n(r) = E_n\psi_n(r) \qquad , \qquad (17)$$

whose solutions $E_n$ and $\psi_n(r)$ are assumed to be obtainable. The $\psi_n(r)$ are complete and orthonormal because the operator on the left of (17) is Hermitian. The solutions of equation (8) for static potentials and arbitrary $\chi$ can then be obtained from equation (10) to be

$$\Psi_{\chi,n}(r,t) = \psi_n(r) \exp[i\{e\chi(r,t) - E_n t/\hbar\}] \qquad . \qquad (18)$$

It can be verified by substitution or from inspection of (10) that equation (18) is a solution of equation (8) with the time independent potentials $A^0$ and $\phi^0$. The $\Psi_{\chi,n}(r,t)$ are solutions of the time-dependent Schrodinger equation with arbitrary gauge for time independent fields. Because of the orthonormality of the $\psi_n(r)$ the $\Psi_{\chi,n}$ remain orthonormal with

$$\int \Psi^*_{\chi,m}(r,t)\Psi_{\chi,n}(r,t)dr = \delta_{m,n} \qquad . \qquad (19)$$

If the Hamiltonian of equation (7) is the exact Hamiltonian operator the solutions $\Psi_{\chi,n}$ are exact; if it is only part of the total Hamiltonian then the solutions (18) form a basis set with which perturbation theory may be carried out.

### 4. Matrix elements
It can be seen from equation (10) that for an operator $V$ to have matrix elements that are independent of gauge and so be physically observable it must satisfy the relation

$$V_\chi = \exp(ie\chi/\hbar) V_0 \exp(-ie\chi/\hbar) \qquad . \qquad (20)$$

Any operator that represents a physically observable quantity must be gauge invariant in this way. Neither the operators $A$ or $p$ are gauge invariant but the operator $p - eA$ is. From equation (9) any operator consisting of positive powers of the components of $(p - eA)$, such as the spin orbit interaction (Frohlich and Studer 1993), will satisfy equation (20). The quantum mechanical operator identified with the particle velocity is $v = dr/dt = [r, H]/i\hbar$ where $H$ is the Hamiltonian. By commuting $r$ with the non-relativistic Hamiltonian $H = (p - eA)^2/2m + e\phi$ the velocity operator $v$ is given by $(p - eA)/m$, so for this Hamiltonian $(p - eA)$ represents $mv$, the mass times the velocity.

The semi-classical Hamiltonian of equation (7) with $p_\chi$ replaced by $-i\hbar\nabla$ obeys the relation

$$H_\chi = \exp(ie\chi/\hbar) H_0 \exp(-ie\chi/\hbar) - e\partial\chi/\partial t \qquad , \qquad (21)$$

and so is not gauge invariant in the above sense. The first term on the right of equation (7), the kinetic energy term, does on its own, satisfy equation (20), but because of the presence of the $\partial\chi/\partial t$ term in (21) the full Hamiltonian $H_\chi$ is not gauge invariant. Accordingly, unless $\partial\chi/\partial t = 0$, the Hamiltonian cannot represent the energy which is a physical observable. This is the issue addressed by this paper whose purpose it is to explain how gauge independent physics can arise out of a Hamiltonian that is not gauge invariant in this sense and in particular how probability amplitudes that are essentially independent of gauge can come from a perturbation theory that uses a non-gauge invariant Hamiltonian.

It is, of course, possible to find operators involving the Hamiltonian that are gauge invariant. $H - e\phi = (\mathbf{p} - e\mathbf{A})^2/2m$, the kinetic energy is one; another is $\{H - i\hbar(\partial/\partial t)\}$ but this is not useful as it is identically zero.

## 5. Scalar potential

Since the difficulty we have encountered is associated with the scalar potential we need to ask the question of what is the appropriate gauge to use to describe the scalar potential so that it accurately represents the physical potential. The physical potential at position $\mathbf{r}$ is the work $W(\mathbf{r},t)$ done to bring a charge $e$ from a point $\mathbf{R}$, usually at infinity, to $\mathbf{r}$ or

$$W(\mathbf{r},t) = -e\int_{\mathbf{R}}^{\mathbf{r}} \mathbf{E}(\mathbf{q},t).\mathrm{d}\mathbf{q} \qquad , \qquad (22)$$

where the force $\mathbf{F}$ on the particle is given by $\mathbf{F} = m\mathrm{d}\mathbf{v}/\mathrm{d}t$ from equation (5) and noting that the magnetic field term vanishes identically. We see that the physical potential involves only the gauge invariant field $\mathbf{E}$. There is no gauge arbitrariness in the physical potential apart from the choice of origin $\mathbf{R}$ which just changes $W(\mathbf{r},t)$ by the same amount everywhere, potential differences remaining unchanged.

It is now seen that an appropriate gauge for describing the physical potential is the multipolar gauge, because, from equation (14), and noting that $\mathrm{d}\mathbf{q} = (\mathbf{r} - \mathbf{R})\mathrm{d}u$ where the integral is along the straight line from $\mathbf{R}$ to $\mathbf{r}$.

$$\phi(\mathbf{r},t) = -e\int_{\mathbf{R}}^{\mathbf{r}} \mathbf{E}(\mathbf{q},t).\mathrm{d}\mathbf{q} \qquad . \qquad (23)$$

When stationary states are considered it is necessary for the physical potential to be unique which means that $\phi$ must be independent of path so the line integral of $\mathbf{E}$ around any closed loop must vanish. From equation (3) the loop integral of $\nabla\phi$ is, of necessity zero, in whatever way $\phi$ is defined, because $\phi$ is a scalar field. The loop integral of $\partial\mathbf{A}/\partial t$ converts to a surface integral of $\partial\mathbf{B}/\partial t$ and if this is to be everywhere zero it follows that $\partial\mathbf{B}/\partial t$ itself must be zero everywhere. If we consider stationary states then $\partial\mathbf{E}/\partial t$ must be zero as well.

By using equations (7,10,16) or (21) it follows that

$$H_\chi^0 \, \Psi_{\chi,n}(\mathbf{r},t) = (E_n - e\partial\chi/\partial t)\,\Psi_{\chi,n}(\mathbf{r},t) \qquad , \qquad (24)$$

and it is seen that as long as $\partial \chi / \partial t = 0$ the eigenvalues of the unperturbed Hamiltonian are unchanged by a gauge transformation. Accordingly, we conclude that in semi-classical wave mechanics the only gauge transformations that leave the eigenvalues of the physical energies unchanged are those that do not depend explicitly on time. However, if the gauge function is taken to have the form $\chi(\mathbf{r},t) = f(\mathbf{r}) + g(t)$ then energy *differences* will remain unchanged. Since only energy *differences* are of significance in wave mechanics, in contrast to quantum field theory in which individual energies are significant, particularly when the coupling to gravity is considered, it appears that gauge functions of this more general form are acceptable in wave mechanics. Although up to now we have used gauge functions that depend on both space and time we now realize that those that depend explicitly on time are not allowable except as mentioned above. The use of the Coulomb gauge is acceptable because the gauge function of the transformation to it does not depend on time *explicitly*, although of course, it does depend on time *implicitly* because it involves the implicitly time-dependent vector potential in the original gauge (Panofsky and Philips 1955). It is in regard to the above restriction on the gauge functions that may be used that semi-classical electrodynamics may be said to be not fully gauge invariant.

**6.    Time development of probability amplitudes**

We now take the potentials to have the more complicated form $\mathbf{A} = \mathbf{A}^0(\mathbf{r}) + \mathbf{A}^1(\mathbf{r},t)$ and $\phi = \phi^0(\mathbf{r}) + \phi^1(\mathbf{r},t)$, where $\mathbf{A}^0$ and $\phi^0$ are the time independent potentials describing static fields discussed in the previous sections of the paper and $\mathbf{A}^1$ and $\phi^1$, which are not necessarily small, describe perturbing fields that generally depend on time. In this approach the gauge is first *fixed*, namely specific mathematical forms are prescribed for $\mathbf{A}^0$, $\phi^0$, $\mathbf{A}^1$ and $\phi^1$, for example $\mathbf{A}^0 = -\mathbf{r} \times \mathbf{B}^0/2$, $\phi^0 = -e/(4\pi\varepsilon_0 r)$ etc., and then *unfixed* by adding in the arbitrary gauge function $\chi$ which, as discussed, essentially depends explicitly only on $\mathbf{r}$. It is necessary to show that the physical results obtained do not depend on $\chi(\mathbf{r})$.

By expanding the square in equation (7), namely

$$H_\chi = \{\mathbf{p} - e(\mathbf{A}^0 + \mathbf{A}^1 + \nabla\chi)\}^2/2m + e(\phi^0 + \phi^1 - \partial\chi/\partial t) = H^0_\chi + V_\chi$$

the wave equation (8) may be written formally as

$$\{ H^0_\chi + V_\chi - i\hbar(\partial/\partial t)\} \Psi_\chi(\mathbf{r},t) = 0 \qquad , \qquad (25)$$

where
$$H^0_\chi = \{\mathbf{p} - e(\mathbf{A}^0 + \nabla\chi)\}^2/2m + e(\phi^0 - \partial\chi/\partial t) \qquad (26)$$

and
$$V_\chi = - \mathbf{A}^1 \cdot \{\mathbf{p} - e(\mathbf{A}^0 + \nabla\chi)\}e/m + e^2(\mathbf{A}^1)^2/2m + ie\hbar(\nabla\cdot\mathbf{A}^1)/2m + e\phi^1 \quad , \quad (27)$$

and where the relation $(\mathbf{p}\cdot\mathbf{A}^1 - \mathbf{A}^1\cdot\mathbf{p}) = -i\hbar\nabla\cdot\mathbf{A}^1$ has been used. The operator $H^0_\chi$ has the form of an unperturbed Hamiltonian in the gauge $\chi$ and apart from $\chi$ has no dependence on time. The basis functions arising from it are obtained as described in a previous section. The perturbation $V_\chi$ of equation (27) depends on time through $\mathbf{A}^1$ and $\phi^1$ as well as $\chi$. However the operator $V_\chi$ is gauge invariant and satisfies equation (20) because the operator $\mathbf{p} = -i\hbar\nabla$ generates the $e\nabla\chi$ term from the phase factor $(-ie\chi/\hbar)$. Are the two sets of potentials $(\mathbf{A}^0,\phi^0)$ and $(\mathbf{A}^1,\phi^1)$ considered to each have their own gauge function $\chi^0$ and $\chi^1$? If they are, then since the sum of two arbitrary functions is another arbitrary function, the two may just be added together to give $\chi$ which may be incorporated in equations (26-27) as shown above. The whole of the time independent perturbation series arising from equations (25-27) is independent of gauge. This is because the matrix elements are because $V_\chi$ is gauge invariant and the energy denominators are because they are energy differences.

To obtain an explicit expression for the probability amplitudes the wavefunction is expressed as a linear combination of the $\Psi_{\chi,n}$ according to equation (11) and as a result of the presence of the perturbation $V_\chi$ transitions take place between the basis states causing the $a_{\chi,n}$ to vary with time. By substituting equation (11) into (25) we obtain

$$(H^0_\chi + V_\chi)\Sigma_n a_{\chi,n} \Psi_{\chi,n} = i\hbar \Sigma_n a_{\chi,n}(t)(d/dt)\Psi_{\chi,n} + i\hbar \Sigma_n \Psi_{\chi,n}(d/dt)a_{\chi,n} \quad . \quad (28)$$

Because the $\Psi_{\chi,n}(r,t)$ satisfy equation (8) with the time independent Hamiltonian from which the basis states are obtained, the first terms on the left and right sides of equation (28) cancel. After multiplying the residual from the left with $\Psi_{\chi,m}^*(r,t)$, integrating over $r$ and making use of orthonormality, equation (28) becomes

$$i\hbar \frac{d a_{\chi,m}(t)}{dt} = \sum_n a_{\chi,n}(t) \int dr \Psi_{\chi,m}^*(r,t) V_\chi(r,t) \Psi_{\chi,n}(r,t) \quad . \quad (29)$$

By making use of equations (18) and (20) the integral over $r$ may be carried out to give

$$i\hbar \frac{d a_{\chi,m}(t)}{dt} = \sum_n a_{\chi,n}(t) V_{m,n}(t) \exp\{i(E_m - E_n)t/\hbar\} \quad , \quad (30)$$

where

$$V_{m,n}(t) = \int dr \psi_m^*(r) V_0(r,t) \psi_n(r) \quad . \quad (31)$$

The important point is that $V_{m,n}(t)$ is independent of gauge. Consequently the equations of motion (30) for the $a_{\chi,m}(t)$ are precisely the same for all gauges. Since the $a_{\chi,m}(t)$ have to satisfy the requirement $\sum_n |a_{\chi,n}(t)|^2 = 1$ for the normalization of $\Psi_\chi(r,t)$ to be preserved the only way that the $a_{\chi,m}(t)$ may differ with gauge is by a phase that is independent of gauge function, by $r$ or by $t$, in other words a number or a global phase factor. It follows that there is no loss of generality in taking the probability amplitudes to be independent of gauge namely $a_{\chi,m}(t) = a_{0,m}(t)$. Consequently the procedure that has always been followed to carry out perturbation calculations, namely setting the gauge function to zero, is justified. This is the main result of the paper.

### 7. Why is semi-classical electrodynamics not gauge invariant?

It is a truth universally acknowledged that the measurable predictions of any physical theory must be independent of gauge. This being the case, it is necessary to provide a justification at a fundamental level for the assertion made in this paper that in semi-classical wave mechanics only a restricted set of gauge transformations is allowable.

The issue of the gauge dependence of the quantum Hamiltonian, equation (7) with $p_\chi = -i\hbar\nabla$, has been discussed by Yang (1976) who defined what he called an Energy Operator which consists of the kinetic energy term and a potential energy term which has no gauge properties. However it is difficult to understand the validity of this approach as the potential energy term in the cases that he considers, which involve the binding of an electron to an atom, is a gauge potential and must be treated as a gauge potential as is done in the present paper. On the other hand Park (1990) considers that the inconsistency is due to the non-relativistic nature of the theory in which time appears not as a quantum variable but as a numerical parameter.

In this paper we take a different view. Semi-classical wave mechanics describes the quantum motion of a particle in a prescribed external electromagnetic field. The field affects the motion of the particle but the particle is not supposed to affect the field. The feature of wave mechanics discussed in this paper, that the Hamiltonian is not fully gauge invariant, is a consequence of this asymmetry in the theory. The same is found in the wave mechanics of the single particle Dirac equation (Stewart 1997a, Stewart 1997b).

In other theories of electrodynamics, the symmetry of interaction between particles and fields is preserved. In non-relativistic classical electrodynamics (Schwinger, DeRaad Jr. *et al.* 1998) the Hamiltonian is

$$H = (\mathbf{p} - e\mathbf{A})^2/2m + e\phi + \int d\mathbf{r}\{\varepsilon_0 \mathbf{E}^2/2 + \mathbf{B}^2/2\mu_0 - \rho\phi\} \quad , \tag{32}$$

where $\rho$ is the charge density. The first two terms give the energy of the particle moving in the field; they correspond to the wave mechanical Hamiltonian equation (7). The remaining terms give the energy of the field in the presence of the particle. It is clear that when the two identical terms involving the scalar potential are cancelled out this Hamiltonian (a) is positive definite, (b) does not contain the scalar potential explicitly and (c) is gauge invariant. Therefore the lack of gauge invariance of the wave mechanical Hamiltonian is not necessarily due to its non-relativistic nature.

The situation is similar in quantum electrodynamics. Its Lagrangian four-density in standard notation (Doughty 1980) is

$$\mathcal{L} = i\overline{\psi}\overset{\leftrightarrow}{\partial}\psi/2 - m\overline{\psi}\psi + e\overline{\psi}A_\nu\gamma^\nu\psi - F^{\mu\nu}F_{\mu\nu}/4 \quad , \tag{33}$$

where the parameters and operators now refer to the bare particles. The coupled Euler-Lagrange equations arising from equation (33) are

$$\partial_\nu\partial^\nu A^\mu - \partial^\mu(\partial.A) = -e\overline{\psi}\gamma^\mu\psi \quad \text{and} \quad (i\partial\!\!\!/ - m)\psi = -eA_\nu\gamma^\nu\psi \tag{34}$$

where the equation on the left incorporates the two Maxwell equations involving the sources and the one on the right is the Dirac equation for a particle in an electromagnetic field. The energy-momentum tensor for the coupled Maxwell-Dirac fields is

$$T^{\mu\nu} = \overline{\psi}\gamma^\mu(i\partial^\nu - eA^\nu)\psi + F^{\mu\sigma}F_\sigma{}^\nu + g^{\mu\nu}F^{\lambda\sigma}F_{\lambda\sigma}/4 \tag{35}$$

and the Hamiltonian three-density $\mathcal{H} = T^{00}$ is

$$\mathcal{H}(x^\nu) = \psi^\dagger\{\boldsymbol{\alpha}.(-i\nabla - e\mathbf{A}) + m\beta\}\psi + (\mathbf{B}^2 + \mathbf{E}^2)/2 \tag{36}$$

where $\psi^{\dagger}$ the adjoint of the Dirac field operator $\psi$ which transforms as in equation (10) under a gauge transformation. The operators act on occupation number states that have no gauge properties. The Hamiltonian is the integral of equation (36) over coordinate space and like the Hamiltonian of classical electrodynamics it does not contain the scalar potential explicitly as does the Hamiltonian equation (7) of wave mechanics. It is readily checked that all the equations (33-36) are gauge invariant: they are unchanged by a gauge transformation. The independence of gauge of the *S* matrix, which corresponds to the probability amplitudes of wave mechanics, is also well established. As a consequence, although quantum electrodynamics suffers from serious unresolved difficulties such as the renormalization needed to deal with divergent self-energies, its fundamental equations are all gauge invariant. Wave mechanics consists of using only the second of equations (34) (in this paper the non-relativistic version of it) and as a result the symmetry and consistency of quantum field theory is lost. That the Hamiltonian of wave mechanics for a particle in an electromagnetic field, equation (7), depends on gauge is hardly surprising since it is derived from a Lagrangian, equation (4), that depends on gauge also. However, since a gauge transformation causes the Lagrangian to change by a term that is a total time derivative, namely $e(d\chi/dt)$, the resulting equation of motion, the Schrodinger equation, retains the same form under a gauge transformation. The underlying reason why semi-classical wave mechanics is not a fully gauge invariant theory is that it is not a properly formulated theory of the interaction between particles and fields. It tells only half the story. It has, of course, other shortcomings such as the inability to describe particle creation and destruction at high energies and these shortcomings led to the development of quantum field theory. However it has been shown in this paper that despite the shortcomings of wave mechanics the probability amplitudes and energies that it predicts are independent of gauge providing it is realized that to maintain a correspondence with the physical energy only certain gauge transformations that preserve energy differences are permissible (see also Stewart 2003).

**Acknowledgements**
Dr R Ball and Professor A R P Rau are thanked for helpful discussions.

**Appendix 1**
The most general solution of the canonical commutation relations $[r^i, p_\chi^j] = i\hbar\delta_{i,j}$ and $[p_\chi^i, p_\chi^j] = 0$ is

$$\boldsymbol{p'}_\chi = -i\hbar\boldsymbol{\nabla} + \hbar\boldsymbol{\nabla} m(\boldsymbol{r},t) \qquad \qquad , \qquad (A1)$$

where $m(\boldsymbol{r},t)$ is an arbitrary dimensionless scalar function of $\boldsymbol{r}$ and $t$ (Dirac 1947). Substituting this into (7) gives

$$H'_\chi = \{-i\hbar\boldsymbol{\nabla} + \hbar\boldsymbol{\nabla} m(\boldsymbol{r},t) - e(\boldsymbol{A} + \boldsymbol{\nabla}\chi)\}^2/2m + e(\phi - \partial\chi/\partial t). \qquad (A2)$$

If the wavefunction is taken to be of the form

$$\Psi'_\chi(\boldsymbol{r},t) = \Psi_0(\boldsymbol{r},t) \exp[i\{e\chi(\boldsymbol{r},t)/\hbar - m(\boldsymbol{r},t)\}]$$

$$= \Psi_\chi(\boldsymbol{r},t) \exp\{-im(\boldsymbol{r},t)\} \qquad \qquad , \qquad (A3)$$

it follows that

$$H'_\chi \Psi'_\chi(r,t) = \exp\{-im(r,t)\}H_\chi \Psi_\chi$$
$$= \exp\{-im(r,t)\}[\{-i\hbar\nabla - e(A + \nabla\chi)\}^2/2m + e(\phi - \partial\chi/\partial t)]\Psi_\chi \ . \quad (A4)$$

Accordingly the quantity $m(r,t)$ in (A1) can be absorbed by the phase factor $\exp\{-im(r,t)\}$ in the wavefunction (A3) and so can be ignored because, unlike in the case of the electromagnetic gauge, it does not give rise to an extra factor analogous to $\partial\chi/\partial t$ in the Hamiltonian.

**Appendix 2**

We derive the relations $\phi_2^0(r) = \phi_1^0(r) - g$ and $A_2^0(r) = A_1^0(r) + \nabla f(r)$ between the potentials of equations (13) and (14) and obtain explicit expressions for $f(r)$ and $g$. For simplicity we take $R = 0$. First we deal with the scalar potential. Using $E(y) = -\nabla_y \phi(y)$ where $\nabla_y$ is the three-gradient with respect to $y$, we get

$$E_1(y) = -\frac{1}{4\pi\varepsilon_0}\int dr' \, \rho(r')\nabla_y \frac{1}{|y-r'|} \quad (A5)$$

and so from equation (14)

$$\phi_2^0(y) = \frac{1}{4\pi\varepsilon_0}\int dr' \, \rho(r')\int_0^r dy.\nabla_y \frac{1}{|y-r'|} \quad . \quad (A6)$$

Performing the integration over $y$ and putting in the limits $y = 0$ and $y = r$ we get $\phi_2^0(r) = \phi_1^0(r) - \phi_1^0(0)$, so $g = \phi_1^0(0)$ where

$$\phi_1^0(0) = \frac{1}{4\pi\varepsilon_0}\int \frac{\rho(r')dr'}{|r'|} \quad . \quad (A7)$$

In a similar way

$$A_2^0(r) = \frac{\mu_0}{4\pi}\int dr' \int_0^1 u \, du \, r\times\{J(r')\times E_{ur}\frac{1}{|ur-r'|}\} \quad . \quad (A8)$$

The triple vector product may be expanded as

$$[J(r')\{r.\nabla_{ur}\frac{1}{|ur-r'|}\} - \{J(r').r\}\nabla_{ur}\frac{1}{|ur-r'|}] \quad (A9)$$

to give two terms in (A8). When the relations between derivatives $u\nabla_{ur}h(ur) = \nabla h(ur)$ and $(r.\nabla)h(ur) = u(\partial h(ur)/\partial u)$ are used (Stewart 1999, Stewart 2000) where $h$ is any function of $ur$, in this case $h(ur) = |ur - r'|^{-1}$, the first term in the integrand becomes $u(\partial h/\partial u) = (\partial/\partial u)(uh) - h$ and (A8) is

$$A_2^0(r) = \frac{\mu_0}{4\pi}\int dr' [\frac{J(r')}{|r-r'|} - \int_0^1 du\{\frac{J(r')}{|ur-r'|} + (J(r').r)\nabla_r \frac{1}{|ur-r'|}\}] \quad . \quad (A10)$$

The first term of this is just $A_1^0(r)$. Now consider the gradient with respect to *r* of the scalar function *f(r)* with

$$f(r) = -\frac{\mu_0}{4\pi} \int d\mathbf{r}' \{\mathbf{J}(\mathbf{r}').\mathbf{r}\} \int_0^1 \frac{du}{|u\mathbf{r}-\mathbf{r}'|} \quad . \quad (A11)$$

Noting that $\nabla\{\mathbf{J}(\mathbf{r}').\mathbf{r}\} = \mathbf{J}(\mathbf{r}')$ we get

$$\nabla f(r) = -\frac{\mu_0}{4\pi} \int d\mathbf{r}' \int_0^1 du [\frac{\mathbf{J}(\mathbf{r}')}{|u\mathbf{r}-\mathbf{r}'|} + \{\mathbf{J}(\mathbf{r}').\mathbf{r}\}\nabla\frac{1}{|u\mathbf{r}-\mathbf{r}'|}] \quad . \quad (A12)$$

The two terms in (A12) correspond to the second and third terms in (A10) to give $A_2^0(r) = A_1^0(r) + \nabla f(r)$ as required. Accordingly the gauge function for the transformation is $\chi(r,t) = f(r) + gt$ where *g* and *f* are given by equations (A7) and (A11).

The integral over u in equation (A11) may be done by standard methods to give

$$f(r) = -\frac{\mu_0}{4\pi} \int d\mathbf{r}' \frac{\{\mathbf{J}(\mathbf{r}').\mathbf{r}\}}{r} \ln\{1 + x\} \quad , \quad (A13)$$

where 
$$x = \frac{(1-r/r')}{(1-\cos\theta)}[\{1 + \frac{2r(1-\cos\theta)}{r'(1-r^2/r'^2)}\}^{\frac{1}{2}} - 1] \quad (A14)$$

and *r* and *r'* are the magnitudes of the vectors and *θ* is the angle between them.